**Effective strain manipulation of the antiferromagnetic state of polycrystalline NiO**

A. Barra*[1], A. Ross*[2,3], O. Gomonay[2], L. Baldrati[2], A. Chavez[1], R. Lebrun[2,4], J.D. Schneider[1], P. Shirazi[1], Q. Wang[1], J. Sinova[2,5], G. P. Carman[1], and M. Kläui[2,3,6,◊]

1. *Department of Mechanical and Aerospace Engineering, University of California Los Angeles, Los Angeles, California, 90095, United States*
2. *Institut für Physik, Johannes Gutenberg-Universität Mainz, D-55099, Mainz, Germany*
3. *Graduate School of Excellence Materials Science in Mainz (MAINZ), Staudingerweg 9, 55128, Mainz, Germany*
4. *Unité Mixte de Physique CNRS, Thales, Univ. Paris-Sud, Université Paris-Saclay, Palaiseau 91767, France*
5. *Institute of Physics ASCR, v.v.i., Cukrovarnicka 10, 162 53 Praha 6, Czech Republic*
6. *Center for Quantum Spintronics, Department of Physics, Norwegian University of Science and Technology, NO-7491, Trondheim, Norway.*

◊klaeui@uni-mainz.de

As a candidate material for applications such as magnetic memory, polycrystalline antiferromagnets offer the same robustness to external magnetic fields, THz spin dynamics, and lack of stray field as their single crystalline counterparts, but without the limitation of epitaxial growth and lattice matched substrates. Here, we first report the detection of the average Néel vector orientation in polycrystalline NiO via spin Hall magnetoresistance (SMR). Secondly, by applying strain through a piezo-electric substrate, we reduce the critical magnetic field required to reach a saturation of the SMR signal, indicating a change of the anisotropy. Our results are consistent with polycrystalline NiO exhibiting a positive sign of the in-plane magnetostriction. This method of anisotropy-tuning offers an energy efficient, on-chip alternative to manipulate a polycrystalline antiferromagnet's magnetic state.

Antiferromagnetic insulators (AFMIs) are promising candidates for device applications because of their high frequency spin dynamics, resilience to magnetic fields, and lack of stray fields[1]. However, functionalizing AFMIs requires reliable electrical mechanisms to detect[2–4] and manipulate[5–9] the magnetic state. For detection, it has been shown that the average Néel vector orientation, **n,** in AFMIs can be identified via the spin Hall magnetoresistance (SMR) present in an adjacent heavy metal (HM) layer[2–4,10]. In such a system, the HM's resistance depends on the angle $α$ between **n** and the probing current, resulting in a $\sin^2(α)$ relationship between the two, due to the additional spin current dissipation path offered by the AFMI[3,4,11]. To manipulate ***n***, magnetic fields can be used in low anisotropy AFMIs (by inducing a spin flop), but magnetic fields are not scalable for device applications or applicable for higher anisotropy materials. As an alternative for memory applications, recent success has been reported in controlling ***n*** by passing electrical current through an adjacent HM layer[5–9]. However, due to the large current densities required, this technique does not scale suitably for device applications because of the inverse relationship between energy consumption and resistance in conventional memory tunnel junctions. Furthermore, passing a high charge current often damages the device under test, resulting in parasitic, non-magnetic contributions to the resistance and short device lifetimes[5,7]. Consequently, the search for alternative electrical approaches that are low power, stable, and scalable are needed such as voltage-based control that is becoming increasingly examined[12–15].

One promising option is to use the large magnetostriction exhibited by many antiferromagnets to control ***n*** through voltage induced strain[13,14,16–19]. Studies have demonstrated the energy efficiency of this method for ferromagnets[12,20], making it ideal to transfer to antiferromagnets[14,18,19], where the magnetostriction can also be sizeable[21–28]. An AFM of general interest in antiferromagnetic spintronics is crystalline NiO, which benefits from both a large intrinsic magnetostriction, originating from spin orbit coupling ($u_s \sim 9 \times 10^{-5}$)[21,24,27,28], and a large SMR response[2–4]. However, to functionalize NiO in a way that leverages its magnetostriction, it would be necessary to grow NiO on piezoelectric materials[13]. One constraint of this approach is the NiO films would most likely be polycrystalline[29–32]. The polycrystalline form might create two undesirable effects, neither of which have been deeply studied. The first is that the macroscopic magnetostriction (measured across an entire sample) can decrease in polycrystalline materials[33]. This could be the case for polycrystalline NiO, where the macroscopic magnetostriction has been reported to be $u_s \sim 1.3 \times 10^{-6}$ in low magnetic fields[34]. However, since macroscopic measurements do not probe the atomic-level structure and anisotropy, polycrystalline NiO's intrinsic magnetostriction might be higher[27,28]. If the intrinsic magnetostriction is too small, polycrystalline NiO may generate insufficient strain-induced anisotropy to adequately control ***n***. This issue was recently addressed in a study which demonstrated that the intrinsic magnetostrictive properties of polycrystalline and monocrystalline NiO can be similar, with macroscopic differences arising due to external factors[32]. The second potential problem related to polycrystallinity arises from the fact that the SMR depends on the average orientation of ***n***. It could be the case that, if measured over sufficient distance, a polycrystalline AFMI may exhibit such a large distribution of *n* that observable SMR signal is unachievable.

In this work, we tackle these key open questions surrounding the strain manipulation of *n*, and the SMR-based detection of average *n*, in polycrystalline AFMIs. This work shows that introducing strain into the antiferromagnetic order of polycrystalline NiO acts in a similar manner to the application of a magnetic field. As the strain applied to the NiO's increases, the magnetic fields required to saturate the SMR response decreases, indicating strain is an effective parameter to manipulate the Néel vector by tailoring the anisotropy.

To examine the strain response of polycrystalline NiO via SMR, a heterostructure of Pt (30 nm)/NiO (7 nm)/Pt (3 nm) along with a Ti (5)/Au (60) adhesion layer was prepared on a [011]-cut single crystal piezoelectric $PbMg_{1/3}Nb_{2/3}O_3$-$PbTiO_3$ (PMN-PT) substrate. The resistance of the bonded device is ~500 Ω, indicating that a majority of current flows in the top 3-nm Pt layer. It has been shown that the contribution to the SMR when the thickness of the metallic layer exceeds the spin-diffusion length rapidly decreases[35]. Any contribution to the SMR from the bottom Pt layer would then be small and with the opposite sign to the top layer, and thus we can attribute the net signal seen later (Fig. 1b and Fig 2) to the SMR response of the 3-nm Pt layer in contact with the NiO. Each layer was deposited via electron-beam evaporation at a base pressure of $10^{-6}$ Torr, following a process that is known to grow polycrystalline NiO free from $Ni_2O_3$ or $Ni(OH)_2$ defects[31].

The thin films were patterned into Hall bars by lithographic methods (see Supplemental), and surrounded laterally by two rectangular Au contacts for applying a voltage and inducing a strain in the PMN-PT substrate, as shown schematically in Fig. 1a. A probing current $j_c$ was passed while measuring a longitudinal voltage *V*, which was used to calculate a resistance $R_{XX}$ (Fig. 1a). All measurements were performed at 200 K, far below the Néel temperature (525 K) of NiO[36]. We chose this temperature to minimize the environmental thermal fluctuations of our experimental system that can mask the small longitudinal SMR signal whilst enabling a full measurement sequence.

Our analysis begins by estimating the biaxial strain applied to the NiO by the PMN-PT substrate using a piezoelectric finite element simulation (Inset of Fig. 1a). The application of a voltage to the Au electrodes (Fig. 1a) stresses the PMN-PT substrate, generating in-plane tensile stress between the electrodes[37]. From our simulations, we determine the corresponding strain developed in the region covered by the Hall bar (dashed outline). The strain in this region is transduced through the basal Pt layer and into the NiO layer, where magnetostriction occurs.

Prior work has established that these stresses transfer efficiently across the basal layers, meaning the transfer does not reduce the stress realized in the device layers[20].

In order to quantify the effect of strain, we first need to establish that SMR can detect the average orientation of *n* in a polycrystalline AFMI. Fig. 1b shows the angular-dependent SMR in polycrystalline NiO/Pt in the absence of external stress. The magnetic field **H** is rotated within the sample plane through an angle $\alpha$, where $\alpha=0°$ occurs for **H**||**j**$_c$. The SMR ratio was calculated by $(R_{XX} - R_{XX}(0))/R_{XX}(0)$ where $R_{XX}(0)$ is the resistance in the absence of a magnetic field. Each dataset was corrected for linear thermal drift when present. Both datasets in Fig. 1b exhibit a periodic $\alpha$-dependent change in resistance, with low resistance for $H \perp j_c$ and high resistance when ***H***||***j*$_c$**. This sign of the signal is surprising for an antiferromagnet[3,4] as it is consistent with the presence of a ferromagnetic component. Positive SMR has been previously reported for antiferromagnets, and is attributable to the existence of some residual interfacial moments[38–40]. We used space-averaged dichroism measurements to exclude the presence of a ferromagnetic interface caused by either proximity-induced magnetism in the Pt layer or unintentional uncompensated Ni monolayer inclusions. However, due to the polycrystalline nature of our films, uncompensated moments may appear at grain boundaries and sample surfaces due to the layered antiferromagnetic order of NiO, leading to a possible defect-induced net magnetization. This net magnetization at grain boundaries can then exchange couple perpendicular to *n*, resulting in a positive observed SMR, as we discuss below. In general, the measured SMR response is sensitive to both a net magnetization *m* (with $\cos^2\alpha$ dependence) and the Néel order *n* (with $\sin^2\alpha$ dependence)[38–40]. The net response is the sum of both, meaning that one observes the angular dependence of the dominating mechanism. Given the magnitude of the critical magnetic fields reported in later measurements (>10 T, Fig. 2), our SMR is clearly dominated by the orientation of *n*, despite the SMR's sign from the exchange coupled moment. Given this, we posit that the SMR is sensitive to *n*, and that the applied strain only affects the magnetocrystalline anisotropy associated with *n*. We thus focus our analysis on the strain-induced effects exhibited by the Néel order, as evidenced in an inverted response study, where we apply strain and measure SMR, eventually corroborating, at the end of the manuscript, that our assumptions about *m* and *n* can explain the experimental observations. We also note that beyond the SMR there may be additional contributions to the electricalsignal from the ordinary magnetoresistance[3,41,42], but such contributions would show up as a $\sin^2\alpha$ dependence under a rotated field[3,41], reducing the net signal we observe. However, the net response we observe, as well as the presence of a magnetic saturation in later measurements, confirms that our signal is dominated by the SMR.

To understand the magnetic field and strain dependence of the SMR, we first provide a theoretical model that physically underpins the experimental results. We assume that the polycrystalline nature of the NiO films leads to an equiprobable distribution of the crystallographic axis in crystallites that are too small to facilitate domain formation. These crystallites act as non-interacting single-domain particles due to the lack of long-range dipole-dipole interactions (see Supplemental). NiO is known to form two types of antiferromagnetic domain, T-domains from contractions of the crystal lattice along the <111> directions and S-domains where the spins align along one of the <112> directions within each T-domain[43–45]. In our model, we consider that the application of strain allows us to change between S-domains within a single T-domain but not to change between different T-domains.

Based on these assumptions, the antiferromagnetic component of the SMR signal is modeled, using standard approaches[3,4] as,

$$\text{SMR} = \rho[1 - \langle (\boldsymbol{n} \cdot \boldsymbol{e}_Z \times \boldsymbol{j_c})^2 \rangle], \quad (1)$$

where $\rho$ is the signal amplitude, $\boldsymbol{e}_Z$ represents a unit vector along the film normal, and $\langle ... \rangle$ indicates the average over all arbitrary orientations of the crystallite axes. The equilibrium orientation of $\boldsymbol{n}$ within each crystallite $\boldsymbol{n}(\mathbf{H})$ relies on the minimization of the energy density $w_{mag}$, which includes three contributions; the magnetic, magneto-elastic and elastic energies (for details, see Supplemental). The additional strain introduced by the piezo-electric substrate leads to a modification of the magneto-elastic and elastic contributions, while the external magnetic field is included only in the magnetic energy. These terms can be used to derive an effective expression for the magnetic energy of NiO given both an external field and stress as,

$$w_{mag} = \frac{1}{2} H_\parallel M_s n_z^2 - H_\perp M_s (n_x^4 + n_y^4) + \frac{M_s}{2} \left( \frac{H^2}{H_{ex}} + \frac{H_{me} \mu_{piezo}}{\mu_{NiO}} \varepsilon \right) (\boldsymbol{e}_X \cdot \boldsymbol{n})^2, \quad (2)$$

where $H_\parallel$ ($H_\perp$) is the out-of-plane (in-plane) magnetic anisotropy of the NiO with respect to the magnetic easy-plane, $M_s/2$ is the sublattice magnetization, $H_{ex}$ is the effective exchange field, $H_{me}$ is the magnetoelastic constant, $\mu_{NiO}$ ($\mu_{piezo}$) is the shear modulus of the NiO (PMN-PT substrate), and $\boldsymbol{e}_X$ is the unit vector in the $X$ direction. Finally, $\varepsilon$ is the strain (linearly proportional to the external stress) applied electrically by the PMN-PT substrate. Due to easy-plane anisotropy of NiO ($H_\parallel \gg H_\perp$), the field dependence of the SMR is characterized by the spin-flop field[2–4]. In a single domain monocrystalline sample, this field corresponds to the magnetic field that saturates the SMR. A spin-flop occurs in a stress-free easy-plane AFM at the spin-flop field $H_{sf}^{(0)} = 2\sqrt{H_{ex} H_\perp}$, at which point $\boldsymbol{n}$ reorients perpendicular to $\mathbf{H}$ if $\mathbf{H}$ lies within the easy-plane. Due to the polycrystalline nature of our samples, the distribution of grain orientations introduces an angle between the applied field and the easy-axes of each grain. This convolution then leads to a distribution of the effective field required to induce a spin-flop in each grain. Experimentally, we will later show that the spin-flop field and the saturation field of the SMR differ, where $H_{sf}$ is a fit parameter representing the grain anisotropy.

Next, we measure the SMR response for a fixed orientation $\mathbf{H}\|\mathbf{j}_c$, where the SMR signal is maximized (Fig. 1b), in strained samples, and use the model to interpret the results. Fig. 2a shows the SMR in the absence of strain along with the theoretical predictions calculated using Eqn. (1) (blue curve), as well as using a model that assumes a single-crystal multidomain sample (grey curve)[3,4]. For all theoretical models, we utilize an approximated spin-flop field of $\mu_0 H_{sf}^{(0)}$=10.5 T and a measured saturation SMR amplitude of $\rho$=86.6·10$^{-6}$. Both models show a similar parabolic response (SMR∝$H^2$) below $H_{sf}^{(0)}$, but different values of the saturation field. For single-crystal samples, the SMR is expected to saturate at $H = H_{sf}^{(0)}$, but the SMR in our polycrystalline samples reaches only 70% of the maximal value at the same field. Instead, the polycrystalline SMR saturates at a much larger field value (i.e. $H \geq H_{sf}^{(0)}$) a behavior consistent with the sample's Néel vector and residual ferromagnetic moment being exchange-coupled despite the positive sign of the SMR. The data is represented best by the polycrystalline model (blue vs grey curves in Fig. 2a). It should be noted that the value of $H_{sf}^{(0)}$ and SMR amplitude extracted here correlates well with recently reports of epitaxial NiO films, with the exception of the SMR sign being opposite[3,4]. This first key result demonstrates that SMR can be utilized to read the average orientation of $\mathbf{n}$ in polycrystalline AFMIs just as well as in single crystalline samples, further confirming a recent report comparing polycrystalline and epitaxial NiO heterostructures in unstrained samples[46]. From a device applications perspective, the significant SMR response in the absence of epitaxial crystalline order opens a substantial number of new material options of technological relevance which warrant further research.

To observe how the SMR response changes under the influence of strain, we measure the SMR with 0 to -200 V applied across the PMN-PT in -50 V steps (Fig. 2b-e). Since the PMN-PT was electrically pre-poled in the same direction as straining voltage was applied, the resulting biaxial piezostrains, $\varepsilon_{YY} - \varepsilon_{XX}$, increase linearly with voltage, meaning the sample experiences strain up to ~$2.80 \cdot 10^{-4}$ in steps of $0.7 \cdot 10^{-4}$. Since our samples are polycrystalline, each NiO grain exhibits isotropic behavior in the easy-plane, it can be assumed that only magnetic fields and strains applied within the easy-plane play a role. Stress is electrically applied by the PMN-PT such that tension appears perpendicular to the magnetic field (along the Hall bar short axis, see inset of Fig.1a). For all values of voltage applied across the PMN-PT, the SMR response increases with applied field. Just as for the unstrained sample (Fig. 2a), the strained SMR response follows the quadratic dependence seen for monocrystalline NiO at low fields, and tends to smoothly saturate at high fields, as approximately described by

$$\text{SMR} \approx \rho \begin{cases} \dfrac{H^2}{H_{sf}^2}, & H < 0.5 \cdot H_{sf} \\ 0.5 \cdot \left[1 + tanh\left(\dfrac{H}{H_{sf}} - 0.75\right)\right] & H > 0.5 \cdot H_{sf}, \end{cases} \quad (3)$$

where the fitting parameter, $H_{sf}(\varepsilon)$, depends on the strain introduced via the PMN-PT substrate.

According to Eqn. (2), strain changes the equilibrium orientation of **n** by balancing the applied field and strain-induced anisotropies. To validate this, our experimental data in Fig. 2b-e was compared with the theoretical model (Eqns. (1)-(3)), utilizing the same saturation SMR value as before ($\rho=86.6 \cdot 10^{-6}$), and we find good agreement for all values of applied voltage. The relation between these two control mechanisms is further illustrated in Fig.3, which shows the calculated SMR value from theory (indicated by the color scale) as a function of both the applied magnetic field and voltage across the PMN-PT substrate. The isolines in Fig. 3 represent the experimental data from Fig 2a-2e but we note there is some deviation between these isolines and the calculated values, due to the sizeable experimental error bars and deviations between the experimental data points and fit curves.

From fitting the experimental data, we extract the spin-flop field $H_{sf}(\varepsilon)$ for each straining voltage. We note that $H_{sf}(\varepsilon)$ has a similar magnitude compared to previous reports for NiO[2–4,47], which further supports the use of our model. Fig. 2f shows $H_{sf}(\varepsilon)$ for each straining voltage along with a fit given by (see supplemental for model details),

$$H_{sf}(\varepsilon) = H_{sf}^{(0)} \sqrt{1 - \frac{\mu_{piezo} H_{me}}{4 \mu_{NiO} H_\perp} \varepsilon}. \quad (4)$$

This equation shows that strain modifies the effective anisotropy, which either enhances or diminishes the effect of the applied magnetic field, depending on the signs of both the magnetoelastic constant, $H_{me}$, and the strain, ε. Our experimental results are consistent with a positive sign of the magnetostriction, implying that $H_{me} > 0$; if it were otherwise, $H_{sf}$ would increase with ε. This indicates that the magnetic order of our sample tends to align with tensile strain (i.e. tensile strain has a similar effect to perpendicular magnetic field). This is opposite to the reported sign of magnetostriction in bulk single-crystalline NiO, which is negative along the easy-axes in the easy-plane[48]. However, one should note that the magnetoelastic coupling coefficients in a polycrystalline sample may differ from those of epitaxially grown films and bulk samples due to the pronounced influence of surfaces and grain boundaries. Given the positive magnetostriction constant reported here, it is not possible to exclude the presence of additional physical factors that may influence our sample but are not included in the theoretical model. One such factor may be the strain-induced redistribution of T-domains rather than S-domains, as has been assumed thus far. T-domains may be formed in NiO when the antiferromagnetic ordering is accompanied by a pronounced contraction along the <111> directions, perpendicular to the easy-plane[49,50]. The distribution of these T-domains in a multidomain sample can, similarly to S-domains, be modified by an external stress. In our experiment, the tensile stress would selectively affect the T-domains with an easy-plane perpendicular to both the Hall bar and *H*, resulting in an increase of the longitudinal SMR. However, further justification of this hypothesis, and a full analysis of relative contributions of T- and S-domains, would require an additional study beyond the scope of this paper.

Regardless of the domain type, our model indicates that it would be possible to induce a spin-flop transition with a sufficient strain. Fig. 2f shows that, the condition $H_{sf}(\varepsilon) = 0$ can be achieved by applying a straining voltage of -275 V across the PMN-PT. However, experimental confirmation of this field-strain equivalence is not reported here, because applying sufficiently high voltage was outside our experimental capability which placed an upper limit on the applied strain to a voltage of -200 V. Our model also indicates (from Eq. (4)) that applied strains could achieve in-plane switching of *n* without an applied magnetic field, which warrants further research in this area.

In conclusion, this paper proposes and experimentally examines the feasibility of the electrical control of an antiferromagnet via a multiferroic composite. Strain anisotropy can be used alongside a magnetic field to control the average Néel vector. These results highlight the potential of strain for future electrically-modulated antiferromagnetic memories and other spintronic devices. We demonstrate that SMR can be used to efficiently read the average orientation of the Néel vector in polycrystalline antiferromagnetic NiO. Applying tensile strain perpendicular to the applied magnetic field substantially lowered the magnetic field needed to overcome the intrinsic anisotropy, consistent polycrystalline NiO exhibiting a positive in-plane magnetostriction.

This work also demonstrates that strain control of antiferromagnets for Néel vector switching-operated memory is feasible and may preserve the energy efficiency benefits of strain-mediated multiferroics as observed in ferromagnets. The fact that the large intrinsic magnetostriction of polycrystalline antiferromagnets dominates the response benefits the development of antiferromagnetic memory control, but further research is required to develop low anisotropy magnetostrictive antiferromagnets.

**Supplementary Material**

See supplementary material for details on the theoretical model employed and sample preparation.

**Acknowledgements**

(*A.B. and A.R. contributed to this work equally.) This material is based upon work supported by or in part by the U.S. Army Research Laboratory and the U.S. Army Research Office under Grant No. W911NF-17-0364. A.B., A.C., J.S., P.S., Q.W, and G.C. are also supported by the NSF Nanosystems Engineering Research Center for Translational Applications of Nanoscale Multiferroic Systems under the Cooperative Agreement Grant No. EEC-1160504. A.R. and M.K. acknowledge support from the Graduate School of Excellence Materials Science in Mainz (DFG/GSC 266) and support from the DFG project number 423441604. L.B. acknowledges the European Union's Horizon 2020 research and innovation programme under the Marie Skłodowska-Curie Grant Agreement ARTES number 793159. Authors from Mainz also acknowledge support from both MaHoJeRo (DAAD Spintronics network, project number 57334897), SPIN+X (DFG SFB TRR 173 –268565370 (projects A01, A03, A11, B02, B11, and B12) and KAUST (OSR-2019-CRG8-4048.2). This project has received funding from the European Union's Horizon 2020 research and innovation programme under Grant Agreement No. 863155 (s-Nebula) and No. 860060 (ITN MagnEFi). O.G. and J. Sinova acknowledge the Alexander von Humboldt Foundation, the ERC Synergy Grant SC2 (No. 610115). O.G. acknowledges support from DFG within the project SHARP 397322108. This work was supported by the Max Planck Graduate Center with the Johannes Gutenberg-Universität Mainz (MPGC). M.K. was supported by the Research Council of Norway through its Centres of Excellence funding scheme, project number 262633 "QuSpin".

**Data Availability**

The data that support the findings of this study are available from the corresponding author upon reasonable request.

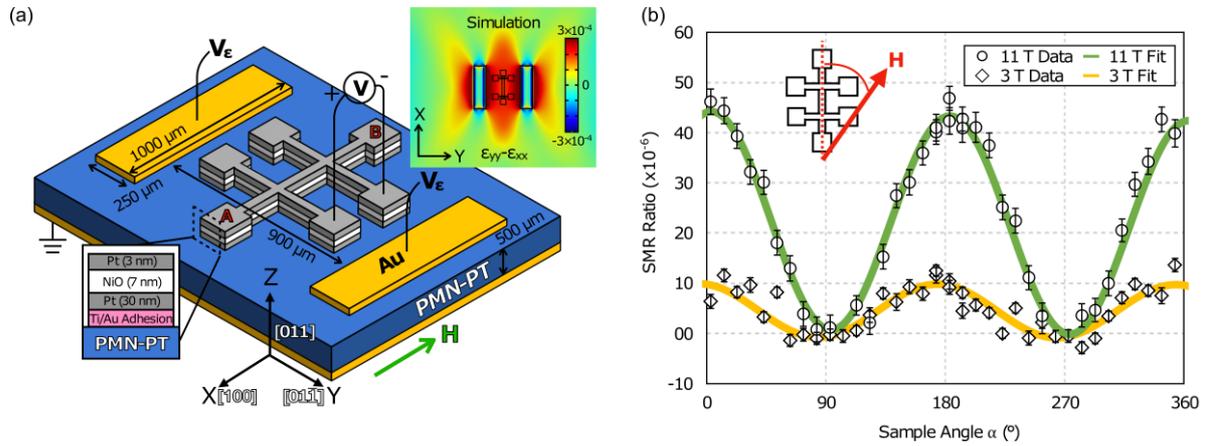

**Figure 1** – Device structure, with the NiO/Pt Hall bar surrounded by patterned straining electrodes (in Au) on either side. A charge current is passed from contact A to B while the longitudinal voltage, V, is measured and used to calculate resistance. Strain is applied by first grounding the back-gate electrode and applying a straining voltage $V_E$ at the Au electrodes. This generates in-plane biaxial tension on the axis perpendicular to the Hall bar's length. Inset: Tensile strain ($\varepsilon_{YY} - \varepsilon_{XX}$), as simulated using COMSOL. The approximate positions of the straining electrodes and Hall bar are shown by black lines. b) Longitudinal SMR as a magnetic field rotates in the sample plane through an angle $\alpha$ (defined between the field **H** and charge current).

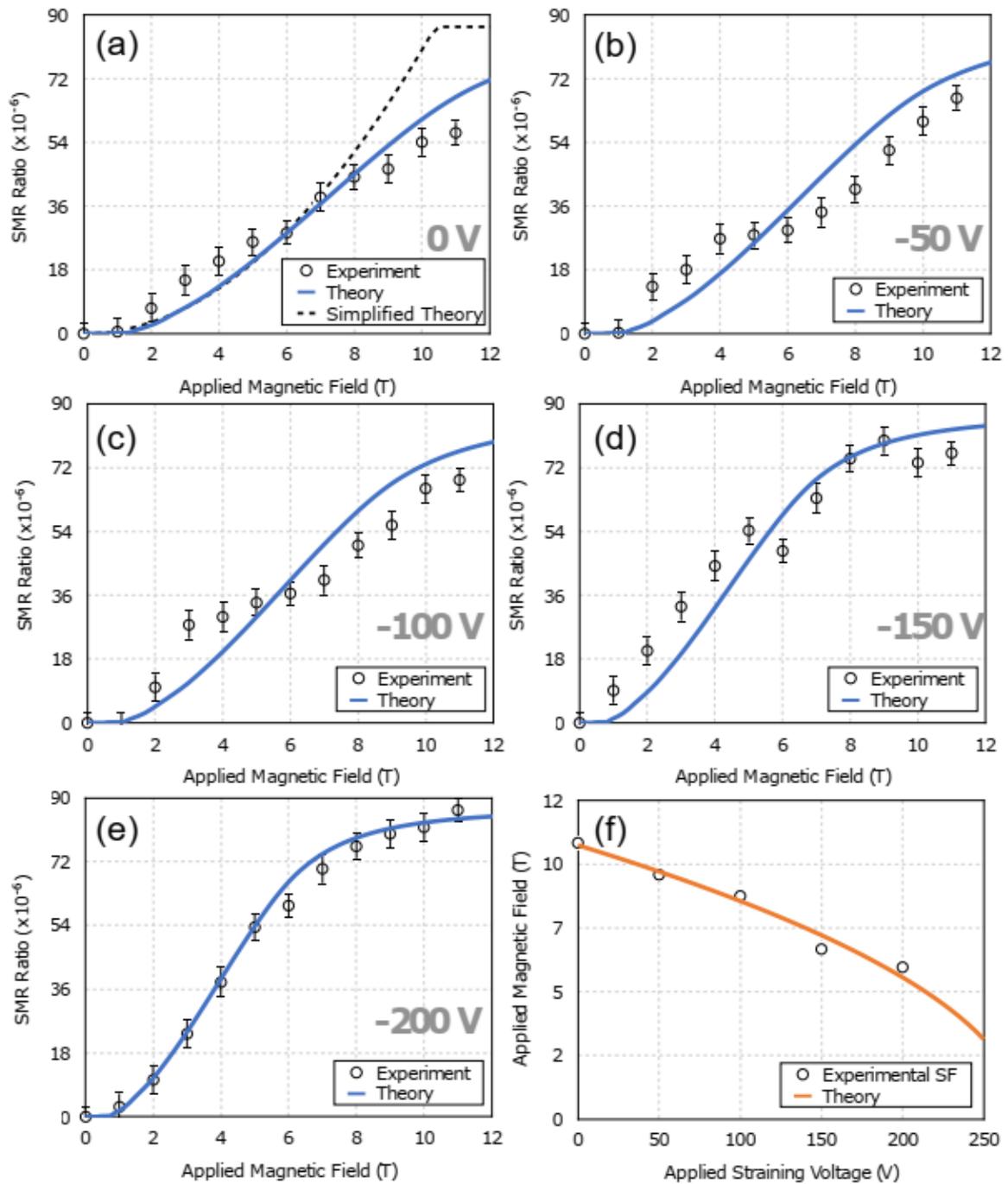

**Figure 2** - a) SMR response as a function of magnetic field in the absence of strain. Theoretical SMR, calculated assuming single-domain and polycrystalline materials, is indicated by the dashed grey and solid blue line, respectively b) – e) SMR response for increasing voltage applied to the PMN-PT, and theoretical fits to the data, based on Eqns. (1)-(3) (blue curves). f) Spin-flop field as a function of applied straining voltage, as extracted from the plots in a)-e). The orange line represents a fit from Eqn. (4).

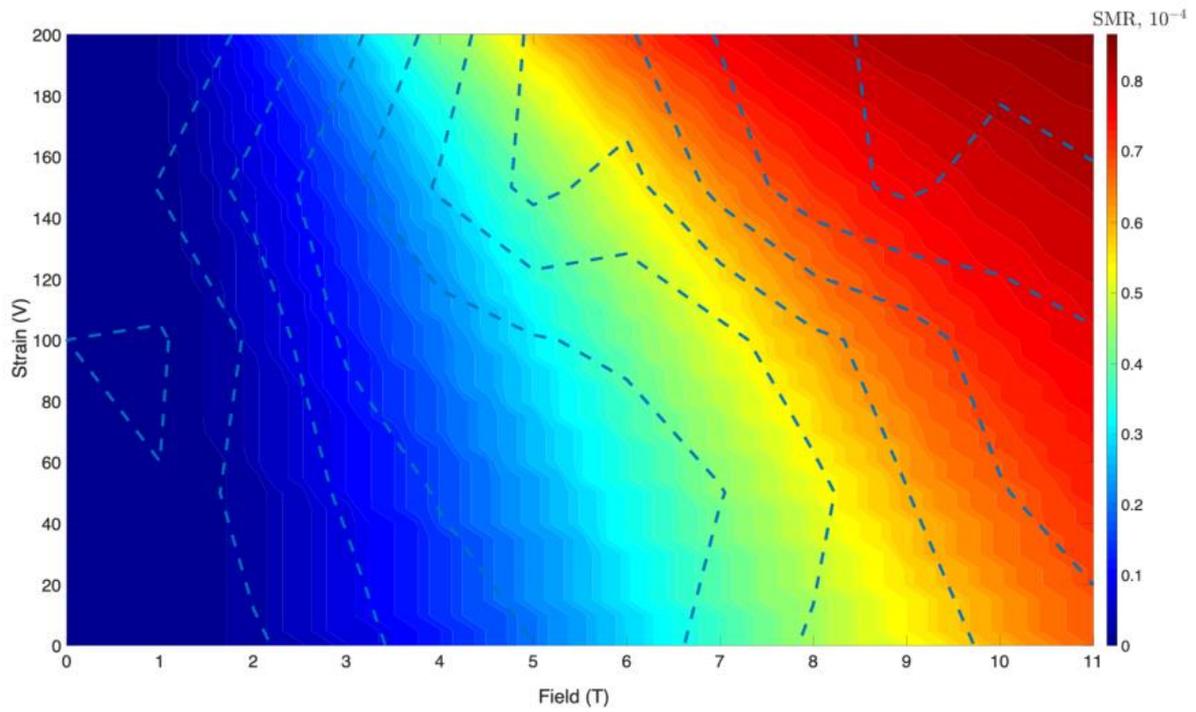

**Figure 3** Strain-field phase diagram of NiO. Color signifies the SMR ratio calculated using Eqns. (1) and (2) with $\rho=86.6\cdot10^{-6}$. Dashed lines correspond to the SMR isolines, which fit the experimental data points (from Fig.2) with the associated error bars omitted to improve the clarity of the figure.